# Synchronized Observations of Bright Points from the Solar Photosphere to Corona

Ehsan Tavabi*

*Physics Department, Payame Noor University (PNU), 19395-3697, Tehran, I. R. of Iran,*
*Institut d'Astrophysique de Paris, UMR 7095, CNRS and UPMC, 98 Bis Bd. Arago, 75014 Paris, France.*

ABSTRACT

One of the most important features in the solar atmosphere is magnetic network and its relationship with the transition region (TR), and coronal brightness. It is important to understand how energy is transported into the corona and how it travels along the magnetic-field lines between deep photosphere and chromosphere through the TR and corona. An excellent proxy for transportation is the *Interface Region Imaging Spectrograph (IRIS)* raster scans and imaging observations in near-ultraviolet (NUV) and far-ultraviolet (FUV) emission channels with high time-spatial resolutions. In this study, we focus on the quiet Sun as observed with *IRIS*. The data with high signal to noise ratio in Si IV, C II and Mg II k lines and with strong emission intensities show a high correlation in TR bright network points.

The results of the *IRIS* intensity maps and dopplergrams are compared with those of Atmospheric Imaging Assembly (AIA) and Helioseismic and Magnetic Imager (HMI) instruments onboard the *Solar Dynamical Observatory (SDO)*. The average network intensity profiles show a strong correlation with AIA coronal channels. Furthermore, we applied simultaneous observations of magnetic network from HMI and found a strong relationship between the network bright points in all levels of the solar atmosphere.

These features in network elements exhibited high doppler velocity regions and large magnetic signatures. A dominative fraction of corona bright points emission, accompanied by the magnetic origins in photosphere, suggest that magnetic-field concentrations in the network rosettes could help couple between inner and outer solar atmosphere.

1 INTRODUCTION

The quiet Sun observed in chromosphere layers is dominated by magnetic bright points (MBPs) or magnetic network at rosettes of supergranule cells (Dunn and Zirker 1973; Tavabi 2014). They may not definitely correspond to the photosphere bright point elements (Muller et al. 2000).

Simon and Leighton (1964) found a strong correlation between chromosphere network magnetic map and quiet Sun supergranulation process by using spectroheliograms, dopplergrams and magnetograms, but the interaction between magnetic field at network (~1 kG) and dynamical flow in the boundary of this giant cells is not theoretically established, and several incompatibilities between supergranulation and the magnetic network have been reported (Snodgrass and Ulrich 1990). Today, this relationship is roughly accepted, suggesting that the magnetic network formation is strongly associated with supergranulation flows (Roudier et al. 2009).

* E-mail: tavabi@iap.fr

The supergranule rosette normally represents several obvious MBPs with high magnetic-field strength while the internetwork bright points are associated with much weaker magnetic fields (Zirin 1993; Keller 1992). The morphology and migration of MBPs can be described as random wandering due to the convection turbulence caused by supergrnular flows from the center to the edge of the cell (Lawrence and Schrijver 1993; Jafarzadeh et al. 2014).

The coronal cells are centered on the enhanced network elements of large-scale regions, and no evidence is found for their relation to internetwork regions. Hagenaar et al. (1997) investigated the cellular patterns of supergranular network, and their model shows the range in outflow strengths is remarkably small at a higher chromosphere. Using cross correlation method between magnetogram and dopp- lergram, Wang and Zirin (1989) confirmed that the supergranule boundaries and magnetic network are roughly correlated and obtained a fine observational support for the relationship between the supergranular magnetic structures and line-of-sight velocities. Coronal bright points (CBPs) have been seen with the help of X-ray and extreme ultraviolet telescopes (Golub et al. 1974;

Habbal and Withbroe 1981). At a large scale, the distribution of CBPs over the solar surface is uniform (Golub et al. 1975) with some deviations from uniformity in latitudes typical for active regions. Golub et al. (1977) found that all X-ray coronal bright points appear in magnetograms as bipolar features, and the separation distance of the bipoles increases with their age. The velocity of the separating motion is in order of 2 kms$^{-1}$, and a typical total magnetic flux is from $10^{19}$ to $10^{20}$ Mx per day of lifetime. The convergence of the bipoles and full cancelation of magnetic elements of opposite polarity have already been reported *(e.g.* see Madjarska et al. 2003). Kayshap and Dwivedi (2017) found that some of the CBPs are formed through the convergence of bipolar magnetic fields, while others appear through a new magnetic field emergence. Koutchmy et al. (1997) provide evidence for the soft X-ray brightening events, which are called coronal flashes. The feature has a short life-time and could be associated with X-ray jets. Sheeley and Golub (1979) found that a CBP consists of several miniature loops (each ~25 Mm in diameter and 12 Mm long) and Alipour and Safari (2015) found two main categories of life-time for CBPs, 52% show typical short life-times of less than 20 minutes and the remaining 48% show long life-times of 6 hr. Identifying and tracking the anchors of the CBP seem to be necessary to understand their sources and physical properties.

A larger number of the HMI magnetic bright points, which are dominated by small scale bright dots, tend to recur at roughly the same location in the photosphere. These characteristics, associated in TR dopplergrams with areas of high line-of-sight velocities, suggest that some of these bright points may result from magnetic reconnections of loops (Tavabi et al. 2015) at source regions in the lower layers of TR and chromosphere. Since the measurement of the coronal magnetic field is almost inaccessible, understanding the real magnetic structure and properties only through the characteristics of the surrounding magnetic structure is indirectly delineated. Filippov et al. (2009) illustrated the geometrical shape of the jets that show a frozen vertical ejection of hot plasma into the higher atmosphere that could introduce rapid heating of corona and mass source into the solar wind flow. One may expect that these CBPs could also be traced throughout the TR and chromosphere given that the photospheric magnetic field lines keep continuously expanding towards the corona. Dopplergrams are the oldest used technique to find supergranulation. The first detection was made by Hart (1954), and the typical length of this structure ranges between around 20 and 30 Mm (Simon and Leighton 1964). We investigate the relationship between the photospheric magnetic-field, chromospheric, and TR dopplergrams in NUV and FUV channels with coronal EUV filtergrams.

## 2 OBSERVATIONS

*AIA/SDO* is an array of four telescopes capturing the images of the Sun's atmosphere out to 1.3R$\odot$ in ten separate EUV wave bands (Title et al. 2006). The images are 4096 x 4096 square with a pixel size of 0"6 in full spatial resolution mode, with a cadence of 12 seconds on 2 July 2014 (Fig. 1) that are considered here. HMI/SDO observes the full disk in the Fe I absorption line at 6173 A with a pixel size of ~0"5, with mechanical shutters and control electronics to take a 45 sec. of line-of-sight magnetic field sequence (Pesnell et al. 2012).

AIA observes simultaneously in the best signal to noise ratio channels in 304 *A* (He II, chromosphere and TR, 50 kK), 171 *A* (Fe IX, TR and corona 6300 kK), and 193 *A* (Fe XII, XXIV, corona and flares, 1.2 MK). *IRIS* 0"3- 0"4 spatial resolution observations, with a pixel size of 0"166 (corresponding to 120 km at disk center) and high cadence, reveal the dynamical behavior of the chromosphere and of TR fine features (De Pontieu et al. 2014). The available high cadence allows the analysis of the evolution of these features and enables us to follow them from the photosphere to the low corona in three channels of different spectral lines, in NUV (2782-2835 Å with the Mg II h and k resonance lines seen in the absorption on-disk and in emission off-disk, similar to the well-known H and K lines of Ca II), and in the FUV ( 1331-1358 Å with the C II line seen in emission), and in 1389-1407 Å (with the Si IV lines seen in emission).

Note that the Mg II lines belong to a low FIP element typical for low temperature plasma situated above the temperature minimum near heights of about 500 km, above $T500$ = 1. The FIP of Mg is 7.65 eV. The C II line is emitted by a high FIP element with excitation potential corresponding to a higher temperature. The Si IV lines are produced at much higher temperatures with an ionization potential of $Si^{+2}$ of 33.5 eV.

Velocity resolution in *IRIS* spectra is 0.5 $kms^{-1}$. In addition, slit-jaw (SJ) images with 166 x 175 *arcsec$^2$* field-of-view (FOV) which reflect off the slit/prism assembly through a filter wheel with four different filters (De Pontieu et al. 2014) are available. This region is located at the center of the solar disk (Fig. 1), where it is quiet, and no active region is observed. It is not a coronal hole region. Some patches of the concentrated magnetic field can, however, exist in FOV.

Fig. 1 provides a synoptic image of the region seen in 171 $A_L$ Fe IX emission from the AIA (*SDO* mission) observations during the *IRIS* observations as reported here.

A very large dense *IRIS* raster over the Sun's disk center was done on 2014 July 2 at 12:12 to 14:00 U.T. with a cadence of 16.3 sec. centered at [49"4; -4"3], the raster covered about 141 x 175 *arcsec$^2$* and took about two hours to complete. Calibrated level 2 data were used; the raster cadence is in order of 16 sec. with a spatial step of 0"35; the 400-pixel long slit covers a 141 x 175 *arcsec$^2$*. The SJ high resolution images were obtained in two channels, showing *i)* the low temperature of TR in C II (1335 $A_L$, logT~3.6 to 4) and *ii)* the much hotter transition region in Si IV (lines near 1400 $A_L$ with logT~4.8); their maximum FOV is also 306 x 175 *arcsec$^2$* (Fig. 4) with a 32 sec. cadence and pixel size is of order of 0"1, including 199 SJ frames for each filters (Tavabi et al. 2015).

## 3 RESULTS

With reference to Figs. 2 up to 6, one could clearly notice that the supergranular cell velocity map is mainly concentrated at the network and much larger line-of-sight velocities and population density placed at the rosettes, as the signal approximately disappears near the center of cells, which is defined as internetwork regions and the flows are dominated by horizontal motions. The bright point intensity fluctuations are considered using the SJI's (1400 $A_L$ and 1335 A) and constructed dopplergrams (Fig. 2) from the dense raster containing spectral information in NUV (Mg II k, 2796 A) resonance line and FUV (C II and Si IV in 1335 $A_L$ and 1400 A) of *IRIS* observations, and AIA and HMI on *SDO* mission in different channel time series. The velocity resolution of dopplergrams is about ±0.5kms$^{-1}$. The dopplergrams are constructed by subtracting red- and blue-wing intensities of spectral lines at fixed offset velocities from the center of the line, *e.g.* ±50kms$^{-1}$. Fig. 3 shows an SJI with faint roundish bright points in the rather hot FUV filtergrams (1400 *A*, T∼80 kK), near the center of FOV, with the entrance slit of spectrograms crossing the region at 13:13 UT. These points are marked by P1, and double points at P2) and are clearly seen as evidence in simultaneous NUV cool spectral (Mg II k, 2796 *A*, T∼10 kK) in the right panel. This region is almost at the center of the solar disk.

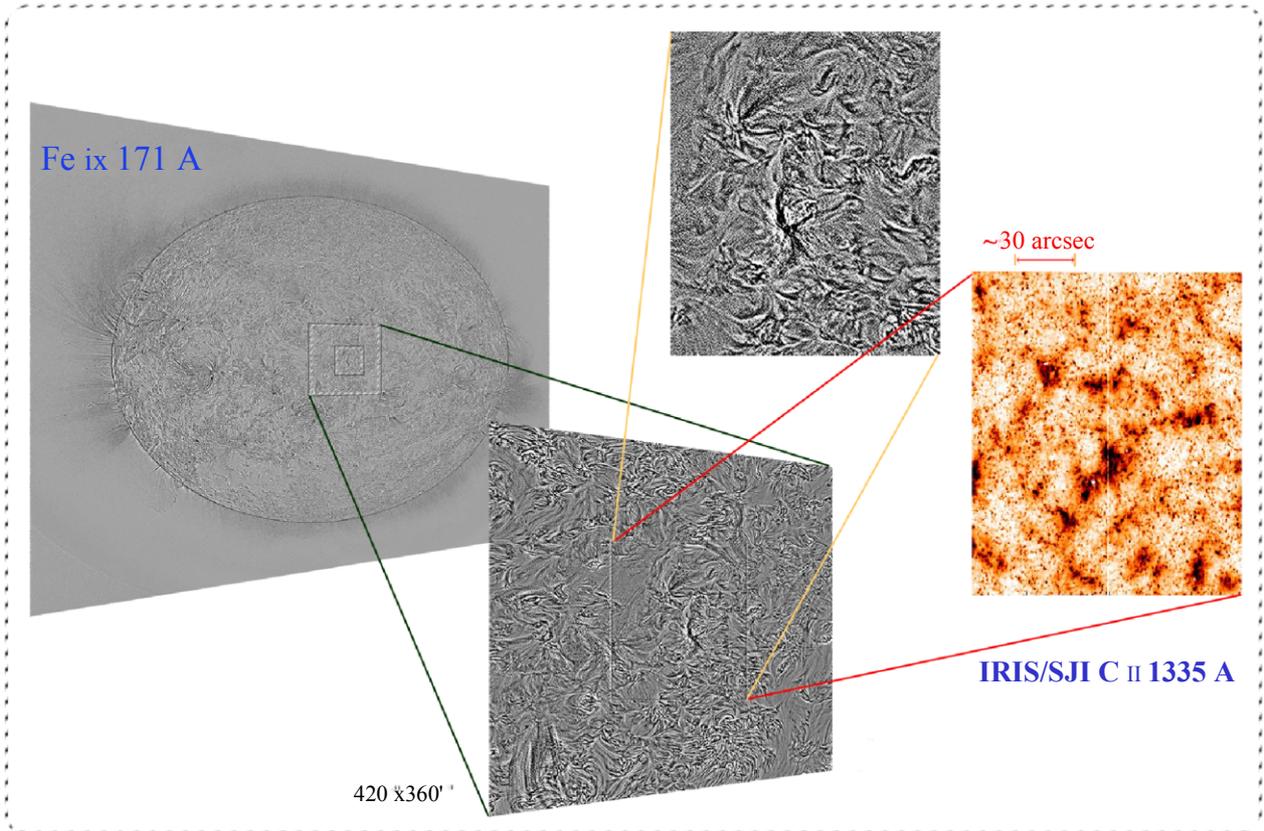

Figure 1. The full disk observation of AIA/SDO in Fe IX 171 A line after summing over 30 min, with 12 sec. cadence in 2014 June 2 beginning at 12:12 U.T. (left panel). The region was observed simultaneously by *IRIS* (right panel) in several TR layers.

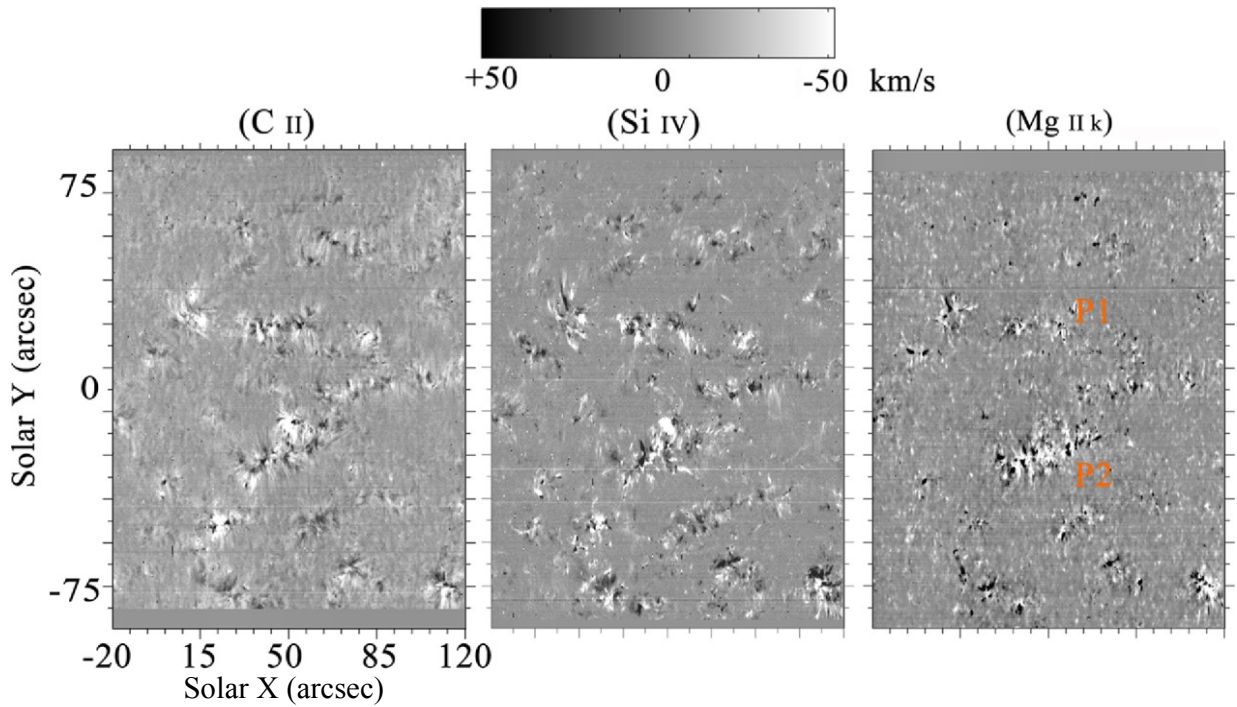

Figure 2. Filtered in the velocity space (white is for blue shifts and black for red shifts). The first is for C II (1335 A), the second for Si IV (1393 A) and the right panel is for Mg II k (2796 A). The two positions with high values of downward velocity are marked by P1 and P2 (Fig. 3). The vertical and horizontal axes indicate the distance in arcsec. The spatial resolutions of raster dopplergrams are adequate, in most cases, to resolve details of the MBPs dynamical aspect.

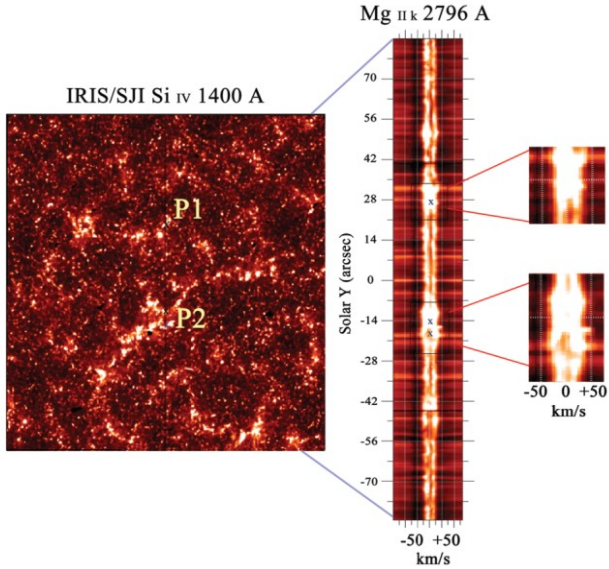

Figure 3. Spatial-temporal properties of bright points. Rapidly evolving bright points are apparent as short-lived with bright features in the blue and red wings (*e.g.,* around ±*50kms⁻¹*) of the chromospheric Mg II k 2796 A spectral line in regions Si IV 1403 *A* slit-jaw image. Highly contrasted and magnified simultaneous spectrograms of the slit position in Mg II k line taken on 02 June 2014 at 13:13:01 UT. The direction perpendicular to the image plane is evident in this line by the large Doppler effects at positions shown by marks (scale at left).

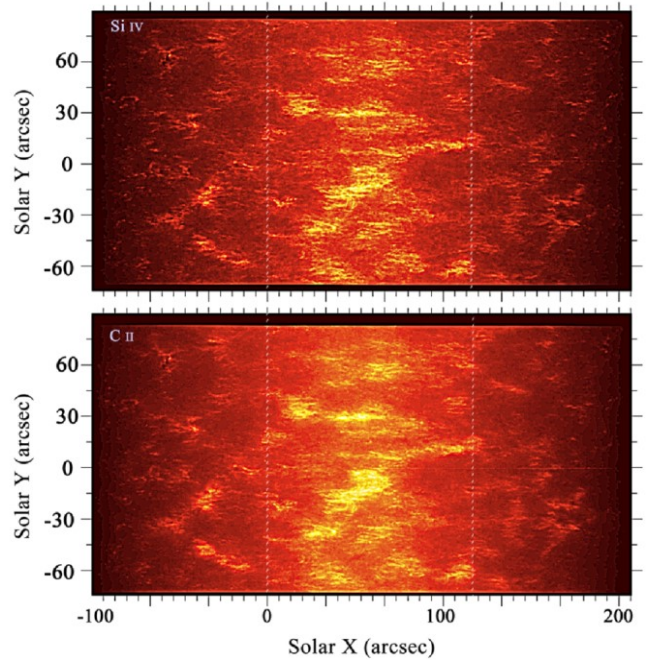

Figure 4. Time-integrated SJI maps of maximum FOV, Si IV 1400 *A* (up) and C II 1335 *A* (bottom) illustrate the moving SJ images (from left to right) in full interval of observation on 2014 July 02 from 12:12 to 14:00 U.T. with a cadence of 16.3 sec. centered at [49"4; -4"3], the SJI maps covered about (∼ 306 x 175 ) and took about two hours to complete. The vertically dotted lines show the FOV of dopplergrams (141 x 175 ).

Fig. 4 shows integrated intensity maps for SJ images in both *IRIS* filtergrams, velocity map FOV in Fig. 2 which is limited between the dotted vertical lines. To illustrate the bright points with higher visibility and decrease the effect of smaller and short life-time brightening, we only observe the prominent region directly related to the TR network. This series was chosen because it is congruent with our primary aim to study MBPs precisely and evaluate a plausible trajectory with height, with respect to photosphere magnetograms, chromospheres, and TR dopplergrams (Figs. 2 to 4). Sheeley and Warren (2012) found this type of cellular features in Fe XII 193 *A* filtergrams of the 1.2 MK corona. This is also confirmed with *IRIS* dopplergram in Si IV line (Figs. 2, 5 & 6 middle panels) that there are mostly downflows in the network, which look reddish in the Doppler map.

Here, the *IRIS* wavelength dopplergrams demonstrate highly correlated dynamical behaviors, and upward (and downward) motions are mainly observed at the boundaries of supergranular cells (Figs. 3 & 4), the correlation coefficient for the dopplergrams in different lines at network regions increases to more than 90% (Fig. 5).

Fig. 6 reveals that most of high-velocity elements coincide with the magnetic network concentrations. The network elements are also close to bright points in the *IRIS* SJ images (Figs. 3, 4 & 5). Fig. 7 illustrates the positions of TR bright points in Si IV (+ marks) and C II (* marks) in the solar quiet region in the same FOV AIA/SDO observation channels as the background snapshots. Physical properties and formation temperature of these two *IRIS* lines closely resemble one another, so these brightenings seem showing the same layer. The shape and size of bright points in both lines are the same with small differences in number. They appear and disappear almost simultaneously (Tavabi et al. 2015; Martinez-Sykora et al. 2015). We integrated the moving SJI's from the beginning to the end of observation series to obtain a maximum FOV (∼ 306" x 175").

In Fig. 4, such time-integrated intensity maps of SJ images are presented, and in Fig. 7, all brightenings of SJI's are compared with the corresponding synthetic AIA filtergrams. AIA images, including the 150 frames in each wavelength channel, have also been integrated over 30 minutes to increase the signal-to-noise ratio.

A close examination of the intensity map reveals the faint presence of small differences at internetwork bright points, while they are in high compliance with the network. By comparing the *IRIS* and AIA filtergrams, we found the counterpart coronal brightenings for all prominent TR bright points in network regions which could be interpreted as a consequence of field aligned plasma motion and give an indirect diagnosis of the trajectory of frozen-in magnetic plasma flows between TR and corona through the layers.

To achieve this purpose, we considered the intensity fluctuation diagrams in details in TR network blightening associated with coronal patches of bright points (Fig. 8). These samples are marked with numbers and show four rather isolated magnetic bright points at network. These points are observed in both *IRIS* SJI's and time-integrated AIA filtergrams over 30 minutes which correspond to this region simultaneously (see Fig. 8).

Almost all coronal bright points can obviously be discerned in space in different wavelengths.

The bright points in 1335 A and 1400 *AL* emission lines seem to come from the same layer, and show an exactly coherent behavior in in-tensity variations with a nice time match (Martinez-Sykora et al. 2015). These marked points as a collection of smaller bright points are associated with cluster counterparts in coronal patches with enhanced emission.

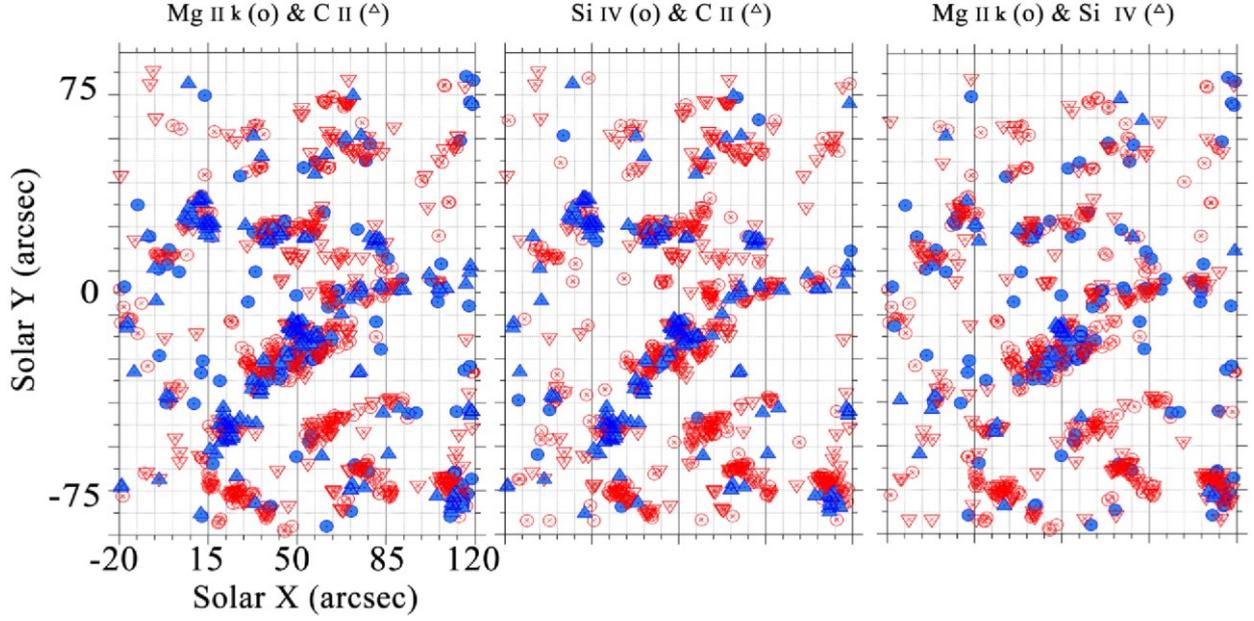

Figure 5. Distribution of high-velocity elements in the studied region in three different *IRIS* lines. Red (blue) symbols with crosses (dots) inside show downward (upward) motions. The correlation coefficient between pairs of distributions reaches > 90% for different *IRIS* lines.

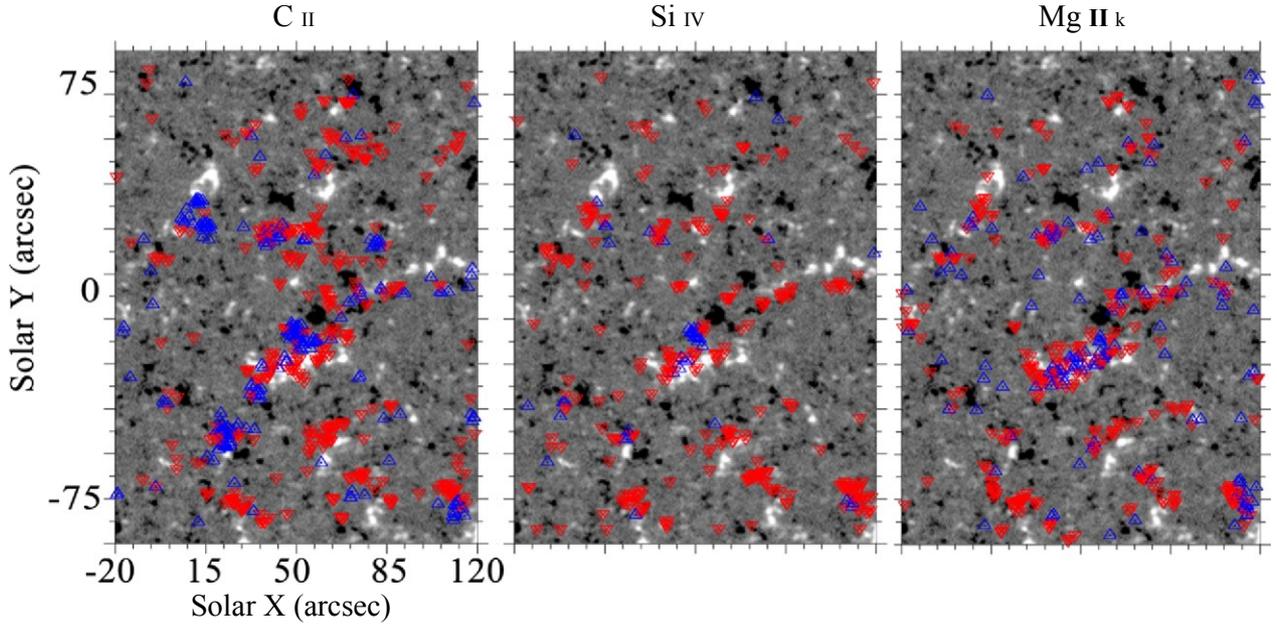

Figure 6. Comparative diagrams show Dopplergrams high velocity position on the HMI/SDO magnetograms for this region in three lines.

To inspect their dynamical behaviors and intensity profiles in detail, we chose these points as they could be well placed on the TR network and are easy to distinguish from the background intensity. These points exist in all lines in TR with a strong intensity variations with short time scales higher than the temporal resolution of the observation cadences.

The variations exhibit a pseudo-oscillatory behavior without the constant frequency in different cases and a regular period of time (Figs. 9 & 10). Therefore, the data seem to suggest that the coronal portion of bright points is heated by some mechanism and then it cools off by radiation and thermal conduction to the TR temperature, where it then turns heated (Habbal and Withbroe 1981).

The correspondence of temporal variations between coronal and *IRIS* data of TR in these bright points is expected to be better for 171 Å and 304 Å lines filtergrams, because temperatures are closer. Then, a better time correlation between corresponding intensity changes is seen in the middle and bottom panels of Figs. 9 & 10. Here, it should be emphasized that the intensity variations are due to an increase in the intensity of the individual pixels after integrating over the area that covers the bright point site. This is not related to a change in their numerical population. Almost all these bright points show the same type of variations in the intensity accompanied with the associated regions in the TR layer which have been plotted synchronic with the SJI/IRIS profiles for the same positions.

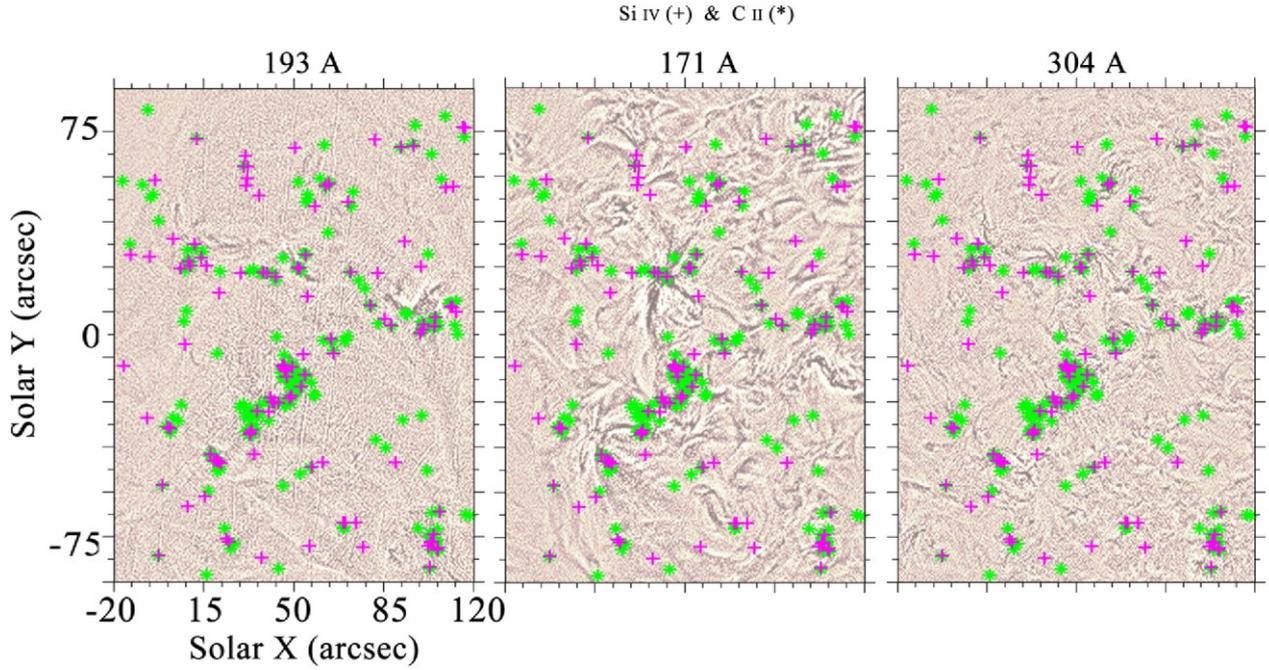

Figure 7. The bright points in transition region SJ images from *IRIS* and AIA/*SDO* rather hot lines show a significant 2D correlation. The background images are the sum results of frames in 30 minutes. After applying the unsharpmask filter in 193 A images, we see long straight lines displaying the readout pixels of the AIA CCD camera.

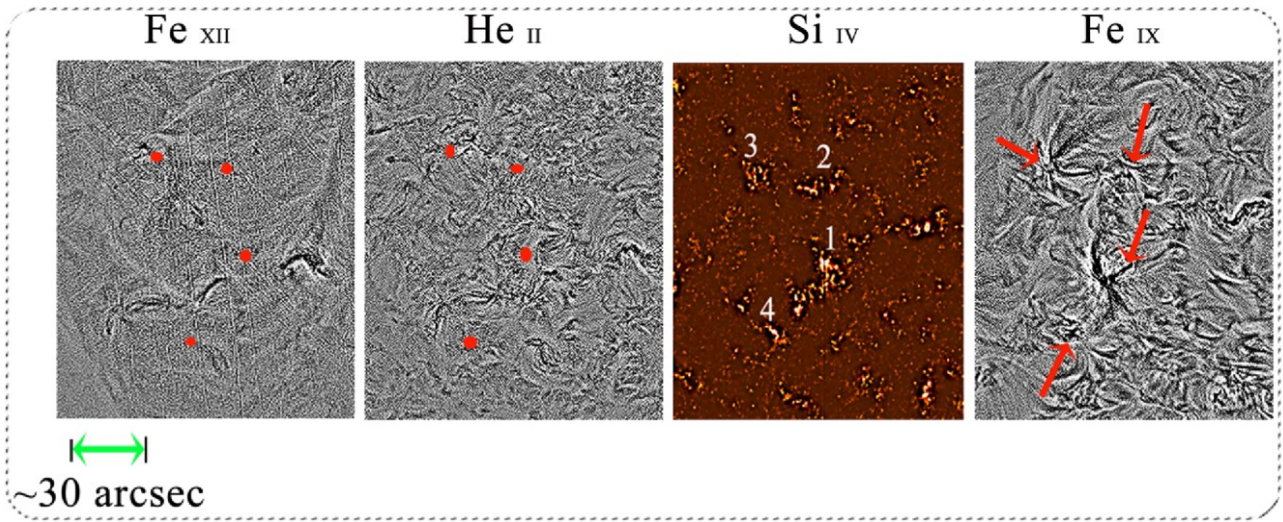

Figure 8. Regions extracted from Fig. 1, displaying the same FOV of *IRIS* observations, some prominent MBP's marked by numbers in Si IV SJI are also marked by arrows in other panels.

**4 CONCLUSIONS**

In this study, we used co-observations from HMI, AIA and *IRIS* instruments that contain photosphere magnetic field, TR, and coronal data, including NUV and FUV spectra of the chromosphere, TR, and coronal filtergrams from the AIA to establish an indirect correlation between bright features observed in the quiet Sun.

HMI magnetograms indicate that field lines are strongly concentrated in photosphere footpoints. Most likely that the field lines expand into the corona.

Our comparative analyzes revealed the strong connection between coronal bright patches and network emission bright points at their bases. Very similar populations of dynamic features appear in the same loci of magnetic bushes with some adjustments. Moreover, the present study illustrates some interesting aspects of MBPs dynamic properties, which have not previously been addressed in the literature.

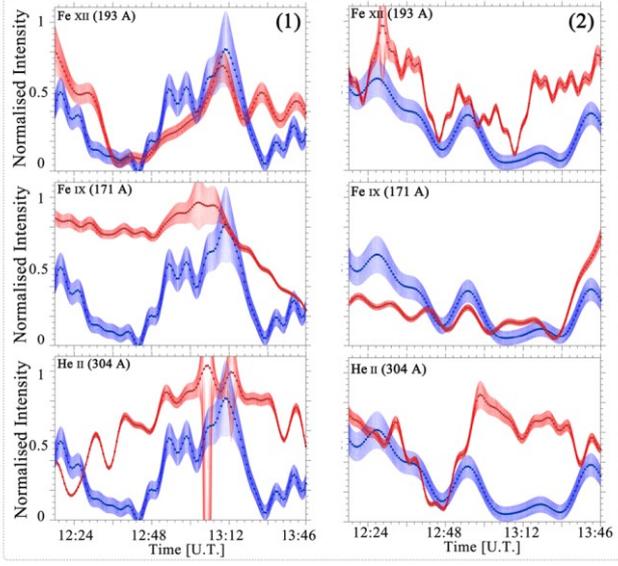

Figure 9. Time-intensity variations in the areal extent (10 x 10 *pixels*$^2$) which is marked as *numbers 1 (left column) and 2 (right column)* in Fig. 8. The common diagram in all panels *(blue squares and confidence bounds)* is for *IRIS* Si IV (1400 A), and simultaneous intensity for AIA channels is illustrated in red.

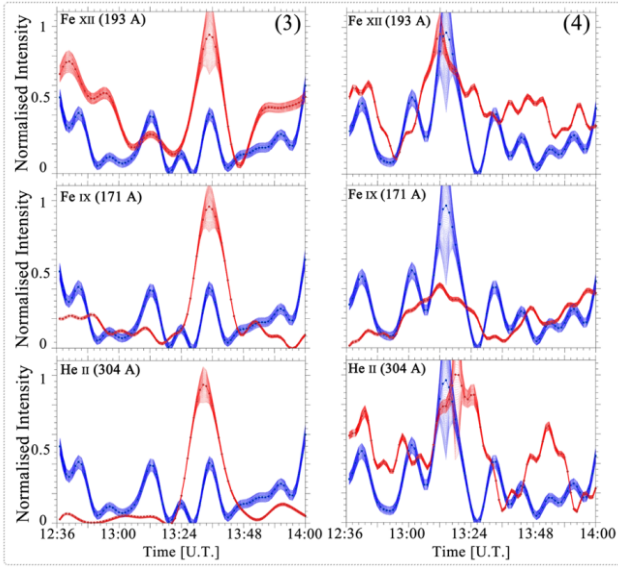

Figure 10. Time-intensity variation in areal extent(10 x 10 *pixels*$^2$) which is marked as *numbers 3 (left column) and 4 (right column)* in Fig. 8. The common diagram in all panels *(blue squares and confidence bounds)* is for *IRIS* Si IV (1400 A), and simultaneous intensity for AIA channels is illustrated in red.

The most important result is that MBPs show very quick variations synchronized with coronal line brightness. High Doppler velocities as well as strong magnetic-field concentrations are very prominent in the chromospheric rosettes.

Sheeley and Golub (1979) reported that the collective coronal bright points are made up of substructures which control the behavior of the chromospheric and coronal emission from these features. Such rapid and simultaneous intensity fluctuations are confirmed in our findings. Prompt and synchronized intensity changes deducible from the radiative and conductive heating and cooling rates are intermittent in the upper solar atmosphere. If this is the case, it can be assumed that, as it is observed and reported, energy supply from TR into the corona could provide sufficient amount of the thermal energy to support the coronal EUV emissions.

Finally, the dopplergrams in cooler lines (C II and Mg II k) demonstrate that a similar range of Doppler velocities in red is always associated with the same values of blue shifted materials at nearby locations. This could be attributed to miniature closed loops in the network regions in the lower layers of TR, whereas hotter Si IV dopplergrams are dominantly reddish at magnetic network.

## ACKNOWLEDGEMENTS

*IRIS* is a NASA small explorer mission developed and operated by LMSAL with mission operations executed at NASA ARC with the contribution ofNSC (Norway). I am indebted to AIA/SDO for providing easy access to the calibrated data and the HMI/SDO team for the high-quality data supplied.

Author are also grateful to the Iran National Foundation INSF.

I extend my gratitude to Prof. S. Koutchmy for his useful discussions, comments and remarks on this study. This work has also been partially supported by the Center for International Scientific Studies and Collaboration (CISSC). The AIA data are courtesy of *SDO* (NASA) and the AIA consortium and the French Institut d'Astrophysique de Paris-CNRS and UPMC that provided a frame for this work.

Comments and remarks from the anonymous referee were very helpful in improving the first version of this manuscript, author also thank for his detailed critical reading of this paper.